# Coulomb explosion sputtering of selectively oxidized Si

P. Karmakar[1*], S. Bhattacharjee[2], V. Naik[1], A. K. Sinha[2], and A. Chakrabarti[1]

[1]*RIB Laboratory, Variable Energy Cyclotron Centre, 1/AF, Bidhannagar, Kolkata 700 064, India*

[2]*UGC-DAE CSR, Kolkata Centre, III / LB -8, Bidhan Nagar, Kolkata 700 098, India*

## Abstract

We have studied multiply charged $Ar^{q+}$ ion induced potential sputtering of a unique system comprising of coexisting Silicon and Silicon oxide surfaces. Such surfaces are produced by oblique angle oxygen ion bombardment on Si(100), where ripple structures are formed and one side of each ripple gets more oxidized. It is observed that higher the potential energy of $Ar^{q+}$ ion, higher the sputtering yield of the non conducting (oxide) side of the ripple as compared to the semiconducting side. The results are explained in terms of Coulomb explosion model where potential sputtering depends on the conductivity of the ion impact sites.

---

*Corresponding author, E-mail: prasantak@veccal.ernet.in

Low energy (~ keV) single and multi charged ion beams induced nanostructure formation has been studied intensively in recent years. Singly charged energetic ions transfer their kinetic energy to the target atoms and create the surface nanostructures following sputtering and diffusion mechanism [1,2]. In contrast, multi charge ions (MCI) carry an internal (i.e. potential) energy corresponds to the sum of the binding energies of the removed $q$ electrons, in addition to the kinetic energy. During the interaction with solid surface the MCI gets back its missing $q$ electrons to become neutralized, which results in hollow atom formation, electron emission, photon emission and potential sputtering [3,4].

The investigation of potential sputtering is one of the most active research areas because such an erosion mechanism is fundamentally interesting as well as important for potential application in defect less cleaning, material selective etching and gentle tool for nanostructuring [5]. A number of investigation have been reported for different materials such as Au(111), HOPG, $CaF_2$(111), LiF(001), $TiO_2$(110), Si(111), $SiO_2$, [6] and ultra thin Pt film [7]. Such investigations are carried out by a range of tools, which include direct sputtering yield measurement by mass loss estimation, secondary ion emission measurement, secondary electron counting, and topographical measurement by scanning probe microscopy. Each of these techniques has their own individual limitations, whether due to thermal fluctuation, charge accumulation, counting statistics, contamination or matrix effect. Nevertheless, the models of potential sputtering say that the interaction of MCI depends highly on the conductivity of target surface [5]. Therefore, it is of great interest to carry out studies pinpointing the effect of potential energy of MCI on a surface having different conductive sectors in the nanometer scale.

In this work, a novel technique has been employed to study the fundamentals of potential sputtering and the effectiveness of such sputtering in gentle tailoring of nanostructures. Spatially resolved semi-conducting and insulating surfaces of nano-ripples are bombarded with multi-charged $Ar^{q+}$ ions. The initial nano-ripple structures are formed by oblique angle single charged oxygen ion bombardment on Si(100) where one side of each ripple is semiconducting and the other side is poor conducting due to preferential oxygen implantation. The dependence of potential sputtering on surface conductivity is shown by comparing the sputtering erosion of coexisting semi-conducting and poor conducting surfaces (which ensures the identical conditions for irradiation and measurement) by topographic and conductivity imaging before and after MCI impact.

Si (100) samples were cleaned with trichloroethelene followed by methanol in an ultrasonic bath. The cleaned and dried Si(100) samples were then transferred in an irradiation chamber for oblique angle oxygen ion irradiation. The samples were first irradiated with 16 keV $O_2^+$ ion beam at $60^0$ angle with respect to the surface normal at a fluence of $2\times10^{18}$ atoms/cm$^2$. The topography and surface conductivity measurements were carried out in air by Scanning Probe Microscopy (Nanoscope IV, Digital Instrument), in contact Atomic Force Microscopy (AFM) and Conducting Atomic Force Microscopy (C-AFM) modes. The samples were again inserted in the irradiation chamber and bombarded with $^{40}Ar^{q+}$ (q = 2, 3, 8, 9) at normal ion incidence. The kinetic energy of the $Ar^{q+}$ was same (32 keV) for all the cases. For MCI irradiation, normal incidence is chosen for symmetric bombardment of both the oxidized and non-oxidized part of the ripple structures. The MCI irradiation is also carried out at grazing incidence ($70^0$) where strong kinetic component of the ion beam is parallel to the ripple direction and only weak component is along normal direction. All the ions were generated and extracted from a 6.4 GHz ECR ion source of the Radioactive Ion Beam Facility at Variable Energy Cyclotron Centre Kolkata [8]. After $Ar^{q+}$ bombardment the nano structures are again imaged by the AFM and C-AFM in air. For C-AFM measurements the bias between the tip and sample was 4 volt.

The ripple formation and selective oxidation both are evident from Fig. 1. It is well established [1] that nano scale ripple structures are formed during the oblique angle ion bombardment on Si (100). Due to the stochastic nature of the incident ions, random roughness is generated on the initial flat surface leading to the development of local curvature [9]. For subsequent ion bombardment at oblique angle, ripple structures are formed on Si(100) because of the competition between curvature dependent sputtering and surface diffusion induced flattening [1,2]. Once the structures are formed, local ion impact angle on the beam facing surface of the ripple is reduced. This results in an increase of implanted oxygen concentration leading to oxidation in the beam facing side of the ripple (Fig. 1). This compositional change causes further reduction of sputtering yield from the front side, and thus the surface of each ripple is decomposed into two phases: a more oxidized portion facing the ion beam and a less oxidized portion on the other side of the ripple. Fig.1 (a1) presents the topographic AFM image whereas Fig. 1(a2) shows the C-AFM current image of the same area which measures the leakage current through the sample for a fixed bias voltage between the tip and the sample. For better comparison, the line profiles of topographic and current images are superimposed in Fig 1 (a3). Fig 1(a3) shows clearly that the leakage current is substantial only at the back side of the ripple, which confirms the coexistence of insulating and semiconducting sectors of the nano ripples.

Homma et al. reported similar structure for 2-10 keV $O_2^+$ bombardment at $45^0$ and Auger mapping showed the asymmetric distribution of implanted oxygen [10]. C-AFM measurement at the bottom of SIMS crater formed by 8 keV $O_2^+$ ion also reveals the same fact [11]. Preferential incorporation of projectile ions in beam facing slope of the ripple is also reported in the cases of 60 keV $Ar^+$ bombarded Si ripple and 16.7 keV $O_2^+$ bombarded Al ripple at the incidence angle of $60^0$ [12,13].

C-AFM measurement of the ripple structures after impact of $Ar^{q+}$ revealed that the oxide area of the ripples is eroded more than the semiconducting part with the increase of charge state of the projectile, though the kinetic energies were the same. Fig.1 (b1-c3) show representative AFM and C-AFM images after $Ar^{3+}$ and $Ar^{8+}$ bombardment. Fig. 1(b1) illustrates the topography of $Ar^{3+}$ bombarded Si ripples whereas Fig. 1(b3) shows the conducting zones of the corresponding ripples. Fig. 1(c2) represents the superposition of the line profiles along the marked lines on topographic and current images. Similarly, Fig. 1 (c1), (c2) and (c3) show the AFM, C-AFM and superposition of line profiles, respectively, of $Ar^{8+}$ bombarded rippled structures. It is clear from Figs. 1 (b1) to (c3) that in case of $Ar^{8+}$ bombardment the area of conducting zones of the ripples are increased compared to $Ar^{3+}$ bombardment although kinetic energies are same in both the cases. In the present experiment the MCI ions carry both kinetic and potential energy as they are not decelerated. Therefore, sputtering yield due to impact of MCI consists of two components, one due to potential sputtering and the other due to kinetic sputtering. To investigate the effect of potential energy, the same kinetic energy ($V \times q = 32$ keV) of the $Ar^{q+}$ ($q = 2, 3, 8, 9$) ions were maintained where V is the extraction voltage and q is the charge state of the ions. Therefore, erosion of ripples due to kinetic energy (32 keV) of $Ar^{3+}$ and $Ar^{8+}$ cases is expected to be similar. In case of $Ar^{3+}$ the stored potential energy is only 84 eV and thus almost equal erosion of both sides of the ripple is expected, however, the increase of conducting area (Fig.1), i.e., preferential erosion of oxide part of the ripple after $Ar^{8+}$ bombardment reveals the effect of potential energy (577 eV), stored in the incident ions.

A large number of C-AFM and AFM images were taken on the $Ar^{q+}$ (q = 2, 3, 8, 9) bombarded ripple structures. A ratio of conducting and non conducting areas of the ripple structures is calculated from each pair of C-AFM and AFM images. The total projected area of the ripple structures are obtained from topographic AFM image, whereas from C-AFM images only the projected area of the conducting part is calculated by WSxM code [14]. The average values of the ratio are plotted as a function of the projectile potential energy, shown in Fig. 2. It is observed that the ratio 'conducting/total' increases and 'non-conducting/ total' decreases with projectile potential energy, which means, higher the potential energy, higher is the erosion of oxide part of the ripples as compared to non oxide part.

The potential sputtering due to impact of MCI has been explained by different existing models [15]. In defect induced desorption model [16], it is assumed that holes and electron hole pairs are created in the valance band of target following the neutralization and relaxation of the MCI ions. In case of alkali halides, the coupling to the lattice is strong leading to trapping of the holes and electron–hole pairs. This leads to the formation of self trapped holes and self trapped excitons. These defects decay further in to color centers which may diffuse to the surface and lead to desorption of target atoms. But defect induced desorption model is shown to be effective only for crystals where a strong electron-phonon coupling is present, e.g. the alkali halides, some oxides and oxidized surfaces.

Inelastic thermal spike model which had originally been developed for Swift Heavy Ion (SHI) induced hillock formation, has been applied for hillocks formation on the MCI irradiated $CaF_2$ by El-Said et al. [17]. During SHI ions bombardment the materials in the ion track is rapidly heated by the electronic energy loss process. If the local temperature exceeds the melting point, the lattice melts and hillocks are created due to relaxation of the internal stress produced by SHI. Although, similar local melting and hillock formation is found by MCI impact, excitation of lattice due to the neutralization of MCI is fundamentally different from excitation mechanism by SHI. This model works well for heavy and very high charge state ion surface interaction.

A number of experimental observations [7,18] have been successfully explained by Coulomb explosion model proposed by Parillis [19]. In this model, when the MCI comes close to the target surface the electrons from the surface fill the high lying Rydberg states of the projectile. The emission of electron from the surface forms charge depletion at the impact point. If the surface is a good conductor, the conduction electrons quickly diminish the charged up domain prior to the explosion, but for poor conductor this charge imbalance will survives for sufficient duration, because of the rather long diffusion length of the electron. Therefore, the situation can be described by the relation between the two time scale $\tau_i$ and $\tau_e$ where $\tau_i$ is the effective time of Auger processes causing the creation of positively charged domain around the multiply charged projectile and $\tau_e$, the time of neutralization of this domain by conducting electrons. For poor conductor $\tau_i << \tau_e$.

The charge domain formed under MCI impact is assumed to be a hemisphere of radius $R_0$. The potential energy 'W' is shared between Coulomb repulsion energy (Ec) and the kinetic energy of the Auger electrons [7,19]. The 'W' could be written as $W \sim R_0^5$ from the energy balance equation [19]. The potential sputtering yield is given by $Y_{ps} = 0.49\pi n(R_0 - a)^3$ , where $a$ is the thickness of the layer from which no particle would succeed to escape during the neutralization time $\tau_e$ [7,19]. Therefore, potential sputtering yield $Y_{ps}$ is proportional to $W^{3/5}$ in the Coulomb explosion model.

The present observation has been compared with the Coulomb explosion model by plotting the ratio of Si to total (Si & $SiO_2$ ) areas with the projectile potential energy, as shown in Fig. 2. The change in the ratio of 'conducting/ total area' with potential energy gives a measure of enhanced sputtering of oxide part of the ripple. A fit (solid line) to the experimental data shows $W^{0.62 \pm 0.09}$ dependence, which is in good agreement with the Coulomb explosion model. Tona et al. [18] studied the secondary ion emission during the impact of $I^{q+}$ on a native $SiO_2$ thin film on Si(111) (an insulator surface) and a clean well defined hydrogen terminated Si(111) (semiconductor) surface. The authors also reported the potential sputtering of ultra thin Pt film in terms of Coulomb explosion model. But in the present case, the direct comparison of sputtering yield of coexisted insulating and semiconducting regions of the nano ripples reduces the difficulties of maintaining the identical experimental conditions and also allows one to estimate the total sputtering by nanometer scale measurement.

Although the kinetic energy of the projectiles was kept constant in all the cases, it is better to study the potential effect by reducing the kinetic energy of the projectiles. To reduce the kinetic component along the surface normal, we bombarded the same ion beam (32 keV $Ar^{q+}$) at grazing ($70^0$) incidence and along the ripple orientation. The observation shows no difference in sputtering between oxide and non-oxide part of the ripples for $Ar^{2+}$ and $Ar^{8+}$ bombardment (data not shown). Peng et al. [20] recently reported sputtering of $SiO_2$ by $Ar^{q+}$/ $Pb^{q+}$ and showed that at larger incident angles, the sputtering yield is dominated by elastic collision between incident ion and material atoms. They also reported that smaller the incident angle, the larger the contribution from the potential sputtering. Therefore, the present observation is consistent with Peng et al. [20].

The present experiment establishes clearly one of the fundamentals of MCI – surface interaction, *i. e.,* the dependence of potential energy erosion on the conductivity of the ion impact site by comparing the sputtering of coexisting oxide and non-oxide surfaces under identical conditions. It shows the capability of multicharge ions in selective etching of surface in nanometer scale and opens up an exciting way for tailoring the shape and dimension of nano structures by ion beam.

The authors thank the members of the RIB group who helped during the ECR operation. The authors also would like to thank Prof D. Ghose for accessing SPM and Mr. S. A. Mollick for assistance during the SPM measurement.

**Figure Captions**

**Figure 1.** (Color online) (a1) AFM and (a2) C-AFM images of Si ripple structures produced by 16 keV $O_2^+$ bombardment at $60^0$, (a3) superposition of one dimensional topographic and current profiles corresponding to the marked lines. After 32 keV $Ar^{3+}$ bombardment at fluence $2\times10^{16}$ ions/ $cm^2$ (b1) AFM topography of the ripple structures; (b2) C-AFM image of the same area showing only the conducting zones; (b3) superposition of topographic (black shed) and current (filled violet) profiles, corresponding to the marked lines. After 32 keV $Ar^{8+}$ bombardment at fluence $2\times10^{16}$ ions/ $cm^2$ (c1) AFM topography of the ripple structures; (c2) C-AFM image of the same area showing only the conducting zones; (c3) superposition of topographic (black shed) and current (filled violet) profiles, corresponding to the marked lines.

**Figure 2.** (Color online) Ratio of semiconducting (Si) to total area, and insulating (oxidized Si) to total area as a function of projectile ($Ar^{q+}$) potential energy.

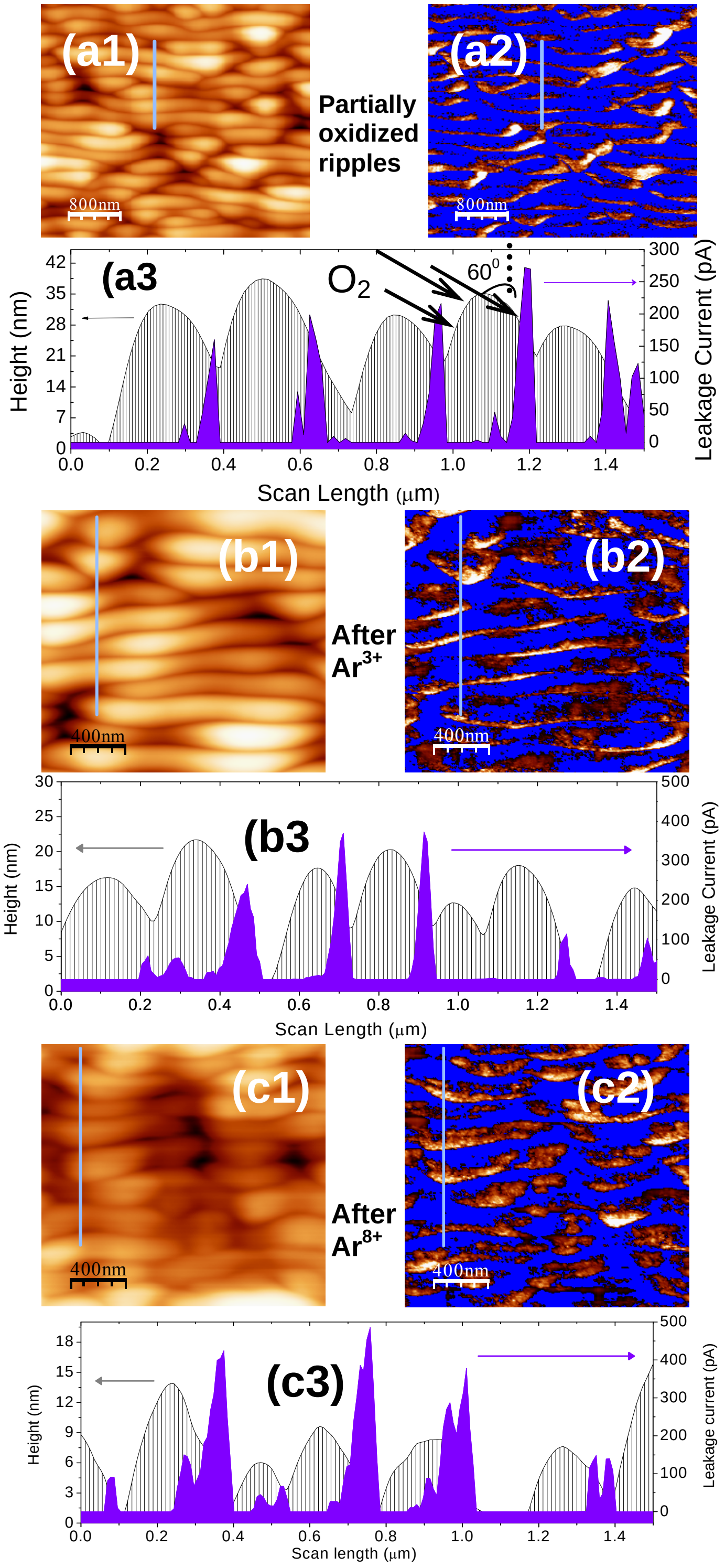

Figure 1

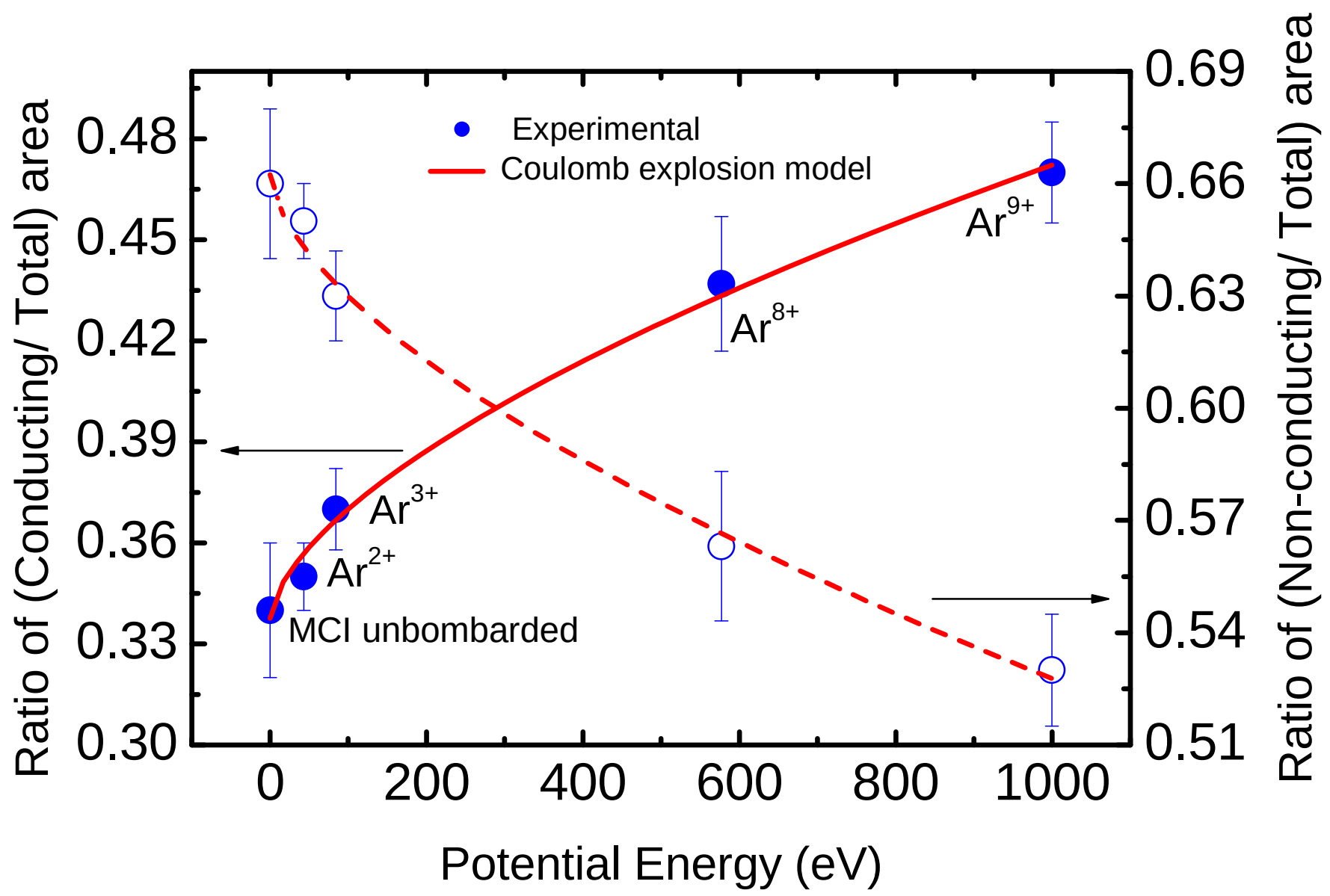

Figure 2